\documentclass{aa}    

\usepackage{graphicx}
\usepackage{txfonts}

\usepackage[colorlinks=true, citecolor=blue, linkcolor=black, urlcolor=black]{hyperref}
\usepackage{natbib}
\usepackage{dcolumn,url,longtable}
\usepackage{amssymb}
\usepackage{color}
\usepackage{tikz}
\usetikzlibrary{shapes,arrows}

\newcolumntype{d}{D{.}{.}{-1}}

\newcommand{\um}{\mu\mathrm{m}}

\begin{document} 

  \title{Shape and spin-state model of tumbling asteroid (319)~Leona}
  \titlerunning{Shape and spin-state model of tumbling asteroid (319)~Leona}

  \author{J. \v{D}urech		   \inst{1}	\and
    J.~L.~Ortiz                 \inst{2}    \and        
    M.~Ferrais                  \inst{3}   \and
    E.~Jehin                    \inst{4}    \and
    F.~Pilcher                  \inst{5}    \and
    J.~Delgado                  \inst{6}    \and
    J.~Hanu\v{s}                \inst{1}    \and
    F.~Marchis                  \inst{7,8}   \and
    J.~L.~Rizos                 \inst{2}    \and
    M.~Kretlow                  \inst{2}    \and
    R.~Iglesias-Marzoa          \inst{9}    \and
    M.~R.~Alarcon               \inst{10,11}    \and
    M.~Serra-Ricart             \inst{10,11,12}    \and
    J.~Licandro                 \inst{11}    \and
    D.~Herald                   \inst{13}   \and
    Z.~Benkhaldoun              \inst{14}   \and
    A.~Marciniak                \inst{15}   
	 }

  \institute{Charles University, Faculty of Mathematics and Physics, Institute of Astronomy, V Hole\v{s}ovi\v{c}k\'ach 2, 180\,00 Prague, Czech Republic, 
             \email{durech@sirrah.troja.mff.cuni.cz} \and
             Instituto de Astrofisica de Andalucia-CSIC, Glorieta de la Astronomia sn, 18008 Granada, Spain \and 
             Florida Space Institute, University of Central Florida, 12354 Research Parkway, Partnership 1 building, Orlando, FL, 32828, USA \and 
             STAR Institute - University of Liège, Allée du 6 Août 19C, B-4000 Liège 1, Belgium \and 
             Organ Mesa Observatory, 4438 Organ Mesa Loop, Las Cruces, NM 88011 USA \and 
             Observatorio Nuevos Horizontes, Camas, Sevilla, Spain \and 
             SETI Institute, Carl Sagan Center, 189 Bernado Avenue, Mountain View CA 94043, USA \and
             Unistellar, 5 all\'e Marcel Leclerc, b\^atiment B, 13008 Marseille, France \and
             Centro de Estudios de F\'isica del Cosmos de Arag\'on, Plaza San Juan 1, 44001, Teruel, Spain \and 
             Light Bridges S.L, Observatorio del Teide. Carretera del Observatorio s/n Guimar, Tenerife, Spain \and 
             Instituto de Astrof\'isica de Canarias (IAC), C/Via Lactea sn, E-38205 La Laguna, Canarias, Spain \and 
             Departamento de Astrof\'isica, Universidad de La Laguna (ULL), E-38206 La Laguna, Canarias, Spain \and 
             Trans-Tasman Occultation Alliance (TTOA), Wellington, New Zealand \and 
             Oukaimeden Observatory, High Energy Physics and Astrophysics Laboratory, Cadi Ayyad University, Marrakech, Morocco \and 
             Astronomical Observatory Institute, Faculty of Physics and Astronomy, Adam Mickiewicz University, S{\l}oneczna 36, 60-286 Pozna\'n, Poland 
	     }

  \date{Received ?; accepted ?}

  \abstract
  {Stellar occultations by asteroids observed from several stations are routinely used to reconstruct the asteroid's sky-plane projections. Together with the asteroid's shape model reconstructed from photometry, such occultations enable us to precisely determine its size and reveal details of its shape. When reducing occultation timings, the usual assumption is that the star has a negligible angular diameter compared to the asteroid, so it is modeled as a point source. The occultation of Betelgeuse ($\alpha$~Orionis) by asteroid (319)~Leona on 12 December 2023 was a rare exception -- the apparent angular diameter of the star was $\sim 50$\, mas, about the same as that of the asteroid.}
  {This work aimed to reconstruct the shape and spin model of asteroid Leona. Then, the projection of that model for the time of the occultation can be computed, which is necessary to interpret the occultation timings and infer valuable information about Betelgeuse itself.}
  {We collected available photometric data of Leona, carried out new observations, and reconstructed a unique convex shape model. Using three other occultations observed in 2023, we scaled this convex model. We also reconstructed an alternative nonconvex model with the same spin parameters and size but showing some surface details that explain better one of the occultations.}
  {We confirmed the tumbling rotation state of Leona and uniquely determined the rotation period $P_\psi = 1172.2 \pm 0.3$\,h and the precession period $P_\phi = 314.27 \pm 0.02$\,h. The volume-equivalent diameter determined by occultations is $59.1 \pm 0.9$\,km. The reconstructed shape model of Leona enabled us to compute its sky-plane projection for the time of the Betelgeuse occultation.}
  {A reliable shape model with accurate dimensions and accurate rotation and precession periods has been reconstructed for slowly tumbling asteroid Leona. It can be used to interpret the observed occultation of Betelgeuse by Leona.}


  \maketitle

\nolinenumbers

  \section{Introduction}

  Stellar occultations by asteroids are routinely observed by both professional and amateur astronomers \citep{Her.ea:20}. These observations are straightforward in that only the times of disappearance and reappearance of the occulted star have to be measured accurately. Together with the observer's geographic position, the timings data are then used to precisely determine the position of the asteroid with respect to the star on the fundamental plane. Nowadays, the scientific importance of such observations is mainly in (i) the precise astrometry of asteroids, which leads to considerable improvement of the orbits and enables better prediction for future occultations, (ii) determining the projected size of an asteroid, and (iii) reconstructing a 3D shape model if there are more chords covering the asteroid's projected silhouette. Occultation observations also led to discoveries of binarity of some asteroids \citep{Gau.ea:22} or the presence of rings \citep{Bra.ea:14}.

  In most cases, the angular size of the occulted star is so tiny compared to the angular size of the asteroid that it can be neglected. So, the star is assumed to be a point source, and the start and end of the occultation are instantaneous when the diffraction effects are negligible. However, some stars have a non-negligible angular size and cannot be treated this simple way. In such cases, the geometry of the occultation is complicated -- the irregular asteroid's projection occults the star's disk that can be non-circular, non-uniformly bright, and have different sizes in different wavelengths. 

  An example of such a star is Betelgeuse ($\alpha$ Orionis), occulted on 12 December 2023 by asteroid (319) Leona. The occultation was not total because the angular diameters of Betelgeuse and Leona were similar. This paper aims to create a shape and spin-state model of slowly tumbling asteroid Leona through light curve inversion and determine an accurate scale of the model by occultations that happened prior to Betelgeuse occultation. Besides providing a highly accurate picture of Leona, this will be critical in interpreting Betelgeuse occultation. This is the first time a shape model of a tumbling asteroid is compared with and scaled by occultations. It is also the first time a nonconvex model of a tumbling asteroid is reconstructed from simultaneous inversion of photometry and occultations.

  \section{Photometric observations of (319) Leona}

    \subsection{Archival data}
    
    Leona's first available photometric light curves were obtained in 2006 by R.~Crippa and F.~Manzini but did not reveal any apparent periodicity.\footnote{\url{http://obswww.unige.ch/~behrend/page_cou.html}} Subsequent observations by \cite{Alk:13c} collected throughout nine nights in April 2013 led to the determination of the period of $14.9 \pm 0.1$\,h. However, the author concluded: ``The flatness of the lightcurve made it difficult to establish an accurate period. This asteroid needs further work.'' 
    
    Then \cite{Pil.ea:17b} carried out new photometric observations for about four months between August and December 2016 with 82 sessions, about 15 min each. They determined the best-fit period to $430.66 \pm 0.18$\,h and realized that Leona's light curve reveals complex rotation; thus, Leona is in an excited rotation state. A Fourier fit with two periods gave the values of $430 \pm 2$\,h and $1084 \pm 10$\,h for the main and the second periods, respectively.

    \subsection{New observations}
    \label{sec:new_observations}

    Motivated by Leona's importance as the body that will occult Betelgeuse, observers carried out new photometric observations to support Leona's shape and spin modeling. We used the data obtained by F.~Pilcher and J.~Delgado available at the Asteroid Lightcurve Photometry Database (ALCDEF)\footnote{\url{https://www.ALCDEF.org}}.
    
    Frederick Pilcher's observations during the 2023/24 oppositions were carried out with a 35-cm, $f/10$, Meade LX200 GPS S-C telescope equipped with an SBIG STL1001E CCD. The telescope is located at Organ Mesa Observatory in Las Cruces, New Mexico, USA, and has the MPC code G50.

    Jesus Delgado's observations from the same period were obtained with a 28-cm, $f/6.85$, Schmidt-Cassegrain equipped with an Atik 414 CCD. The telescope is located in Observatorio Nuevos Horizontes, Camas, Spain, and has the MPC Code Z73.

    Both datasets were processed using the MPO Canopus version 10.7 software. The data were internally calibrated in the R filter, which allowed us to use them as a single light curve (see Fig.~\ref{fig:lcfit}). The internal calibration of individual nights is crucial because single light curves from individual nights are flat due to Leona's long rotation period. This was also why we did not use observations by D.~J.~Higgins, although they are also available at ALCDEF -- they are not internally calibrated, so they carry very little information about Leona's rotation. We reduced the number of data points for modeling purposes described in the next section. Because the brightness changes slowly due to the long period, the original sampling of light curves is too dense, and the large number of data points slows the computation. Therefore, we reduced the amount of data (by exclusion) so that the minimum time between subsequent observations was 3 minutes.

    \paragraph{TRAPPIST} Photometric observations of Leona were obtained from 17 September 2023 to 24 February 2024 with one or both of the TRAPPIST telescopes to sample Leona’s rotation curve each night. We used TRAPPIST-South (TS) located at the ESO La Silla Observatory in Chile \citep{Jehin2011}, and TRAPPIST-North (TN) located at the Ouka\"{i}meden observatory in Morocco. They are 0.6-m Ritchey-Chr\'{e}tien telescopes operating at $f/8$. The camera on TS is a FLI ProLine 3041-BB CCD camera with a 22 arcmin field of view and a pixel scale of 0.64 arcsec/pixel. TN is equipped with an Andor IKONL BEX2 DD camera providing a 20 arcmin field of view and pixel scale of 0.60 arcsec/pixel. We used the blue-cutting Exo filters, no binning, and exposure times of 120\,s. The photometry was obtained with the Photometry Pipeline \citep{Mom:17}, and the magnitudes were calibrated to the Rc Johnson-Cousins band with the PanSTARRS DR1 catalog using typically 100 field stars with solar-like colors in each image.
    
    \paragraph{Javalambre Observatory} Observations of the asteroid were conducted with the Javalambre Observatory (Teruel, Spain) using the 0.4m telescope equipped with a ProLine PL4720 CCD camera. The CCD images were acquired from 23 September till 19 December 2023, twice per night with 3--4-hour separations to ensure clear identification and optimal measurement of its light curve. Imaging was performed without filters (150-s exposures) to enhance detection and approximate Gaia's spectral bandwidth. Standard dark and flat-field calibrations were made. Astrometric calibration used the UCAC4 catalog, and the asteroid's position was determined via Horizons' ephemerides. Only Gaia-matched stars with solar-like color indices ($0.377 < \mathrm{BP}-\mathrm{RP} < 0.983$) were used for calibration. The asteroid's calibrated magnitude was computed using flux ratios between the asteroid and comparison stars. Iterative sigma-clipping excluded outliers to minimize contamination from poorly measured or variable stars. Photometric uncertainty was determined from the dispersion of surviving measurements. Some images were excluded due to issues like contamination by overlapping stars or anomalously low numbers of comparison stars. These omissions had minimal impact on the final light curve analysis.

    \paragraph{Teide Observatory} Photometric observations were conducted using the Two-Meter Twin Telescopes (TTT) facility located at the Teide Observatory on Tenerife Island (Canary Islands, Spain). This facility currently operates its first two telescopes, TTT1 and TTT2, each featuring 0.80-m apertures, altazimuth mounts, and two Nasmyth foci with focal ratios of $f/6.8$ and $f/4.4$, respectively. At the faster focal port, a QHY411M camera equipped with an sCMOS sensor (3.76~$\um$~pixel, 151 megapixels) is installed \citep{Alarcon2023}. The camera provides a full-frame field of view of $52.3\arcmin \times 39.2\arcmin$ with a plate scale of 0.221\arcsec~pixel. Observations were consistently conducted using an \textit{g\arcmin} SDSS filter. The data reduction process followed standard techniques. Images were first corrected for bias and sky flat-field. Subsequently, a $2\times2$ average binning was applied, and a central region measuring $18.4\arcmin\times18.4\arcmin$ was cropped. Aperture photometry was performed using the \textit{Tycho Tracker}\footnote{\url{https://www.tycho-tracker.com}} software \citep{2020Parrott}. For astrometric calibration, the images were aligned using bicubic interpolation, downsampled by a factor of two, and processed with \textit{Astrometry.net} \citep{Lang2010}. Photometry employed a fixed circular aperture with a radius of 2\,$\times$\,FWHM, determined from the first image. The sky background signal was estimated using an outer annular ring placed at 4\,$\times$\,FWHM. The same aperture parameters were applied to the selected comparison stars, constrained by $0.60<(B-V)<0.70$. To minimize positional uncertainties arising from the ephemerides, the start and end points of the track were marked manually.
    
    \paragraph{Unistellar} Leona was also observed in 2023--2024 by more than 40 citizen astronomers centralized in the Unistellar Network \citep{Marchis2020}, who contributed optical data through coordinated light curve observations. The data processing is standard -- it involves subtraction of background noise, World Coordinate System solving, stacking, and performing aperture photometry. Specifically, each 20-minute Leona observation was first stacked into 2-minute stacks. We used aperture photometry to estimate the brightness of Leona for each stack and then took the average of all of the magnitudes as the true magnitude for the observation. Our approach of having a single brightness value for each 20-minute observation is justified by Leona’s long rotation period. While these observations provided valuable coverage, there was a significant scatter in the data. Therefore, we rejected observations where stellar contamination influenced the measured flux (either in the asteroid aperture or the background annulus), which accounted for all of the most apparent outliers. We obtained 105 individual measurements that we consider reliable. However, due to the larger photometric scatter compared to other photometric datasets, Unistellar data were not used in Leona's shape modeling to ensure the highest accuracy of the physical model. Nevertheless, we present the data alongside model predictions for comparison in Fig.~\ref{fig:lcfit}. Unistellar data match the modeled light curve well, even for epochs not covered by other datasets. 
    All citizen astronomers who contributed with data are listed in Table~\ref{tab:unistellar}.
   
    \subsection{Sparse photometry from surveys}

    Leona's light curves described above cover only two apparitions in 2016 and 2023/24. To have more information about Leona's changing brightness under a wider range of viewing and illumination geometries, we also used sparse-in-time photometric data from Gaia DR3 \citep{Tan.ea:23, Bab.ea:23} and ATLAS \citep{Ton.ea:18, Ton.ea:18b, Smi.ea:20, Hei.ea:18}. The Gaia data are available through the Gaia ESA Archive\footnote{\url{https://gea.esac.esa.int/archive/}}, 28 photometric measurements were obtained between 2015 and 2017. ATLAS photometry of Leona was extracted from the ATLAS Solar System Catalog V2\footnote{\url{https://astroportal.ifa.hawaii.edu/atlas/sscat/}}; it consists of about 2850 individual measurements in two filters obtained between 2015 and 2024. We removed ATLAS observations within $10^\circ$ around the galactic plane because background stars might have contaminated them. This reduced the number of data points to about 2000.
  
          \begin{table}[t]
            \caption{Spin parameters of the convex model.}
            \label{tab:spin}
            \begin{tabular}{l c c}
            \hline
             \multicolumn{2}{c}{parameter}   &   value   \\
            \hline
            angular momentum vector     & $\lambda$   & $88 \pm 3\degr$      \\
                                        & $\beta$     & $11 \pm 3\degr$      \\
            Euler angles                & $\phi_0$    & $116 \pm 4\degr$      \\
                                        & $\theta_0$  & $34.8 \pm 0.5\degr$    \\
                                        & $\psi_0$    & $10 \pm 0.3\degr$    \\
            principal moments of inertia & $I_1$       & $0.495$        \\
                                        & $I_2$       & $0.845$        \\
            periods                     & $P_\phi$    & $314.27 \pm 0.023$\,h    \\

                                        & $P_\psi$    & $1172.2 \pm 0.27$\,h     \\
            Hapke's parameters          & $w$         & $0.044$         \\
                                        & $h$         & $0.11$          \\
                                        & $B_0$       & $2.2$          \\
                                        & $g$         & $-0.41$       \\
                                        & $\bar\theta$    & $20\degr$  \\
            geometric albedo            & $p$       & $0.07$          \\
            \hline
            \end{tabular}
            \tablefoot{The Euler angles are given for  JD$_0$ 2456908.0.}
        \end{table}

  \section{Shape and spin model reconstruction}
  \label{sec:shape}

    Convex shape models of asteroids are routinely reconstructed from their photometric data by the light curve inversion method of \cite{Kaa.Tor:01, Kaa.ea:01}. The spin state is described by the direction of the rotation axis, the sidereal rotation period, and the initial orientation of the shape for some chosen epoch. Projecting the model on the fundamental plane for any epoch is simple as long as the rotation period is known accurately. This approach has been used by \cite{Dur.ea:11} or \cite{Mar.ea:23}, for example, to scale shape models reconstructed from light curves by occultations. However, this approach has been applied only to asteroids in principal-axis rotation, which is not the case for Leona.
    
    The tumbling rotation state of Leona was revealed by \cite{Pil.ea:17b} from the complex shape of its photometric light curve. This was confirmed with new observations in 2023/24. The excited rotation state complicates predicting Leona's orientation during Betelgeuse occultation. In the case of a tumbling asteroid, the light curve inversion approach is the same as in the case of relaxed rotation -- the convex shape model and spin state are reconstructed from photometric data. Then the projection can be computed for any epoch. In practice, however, finding a unique solution for a tumbler is more complicated than for a principal-axis rotator because the number of spin parameters is larger. The set of parameters describing the tumbling rotation is $(\vec{L}, \phi_0, \theta_0, \psi_0, I_1, I_2)$ or equivalently $(\lambda, \beta, \phi_0, \psi_0, P_\psi, P_\phi, I_1, I_2)$, where $\lambda$, $\beta$ are ecliptic coordinates of the direction of the angular momentum vector $\vec{L}$, parameters $\phi_0$, $\theta_0$, and $\psi_0$ are initial Euler angles, $P_\psi$, $P_\phi$ are rotation and precession periods, and $I_1, I_2$ are normalized principal axes of the inertia tensor ($I_3 = 1$). The light curve inversion method for free precession asteroids was developed by \cite{Kaa:01}, and it has been successfully applied to asteroids Apophis \citep{Pra.ea:14}, 2008~TC3 \citep{Sch.ea:10}, or Krylov \citep{Lee.ea:17}. The critical aspect is to determine the main periods in the light curve signal. The actual physical frequencies $1/P_\psi$ and $1/P_\phi$ are linear combinations of the light curve frequencies. In practice, all reasonable combinations are usually tested, and the best one corresponding to the correct model is selected. 

    However, in the case of Leona, we used a simplified ellipsoidal model to scan the parameter space efficiently.
        
    \subsection{Ellipsoidal model}
    \label{sec:ellipsoids}
  
        The convex shape model approximation usually employed to model light curves of tumbling asteroids is flexible enough to reproduce light curves of real asteroids with nonconvex shapes. As described by \cite{Kaa.Tor:01}, the optimization is performed in a so-called Gaussian image, i.e., the curvature of its surface describes the shape, and no 3D shape model is available during optimization, only areas and normals of surface facets. Vertices of the final 3D shape model are then computed using the Minkowski procedure \citep{Lam.Kaa:01}. For this reason, moments of inertia $I_1$ and $I_2$, which are free parameters of optimization, are not coupled with the shape; they are independent parameters. When the best-fit model is found and reconstructed with the Minkowski procedure, its inertia tensor is computed (assuming uniform density), and its principal moments $I_1^\mathrm{shape}$ and $I_2^\mathrm{shape}$ are compared with those that were found as kinematic parameters. Ideally, these two sets of moments of inertia should be close to each other for a physically self-consistent model. 
  
        When searching for the best set of kinematic parameters ($P_\phi$, $P_\psi$, $I_1$, $I_2$), an efficient approach is to first fit the light curves with a two-period Fourier series \citep{Pra.ea:05} and then try various linear combinations of main Fourier frequencies as physical frequencies $1/P_\phi$ and $1/P_\psi$. This way, models of 2008~TC3 \citep{Sch.ea:10} or Apophis \citep{Pra.ea:14}, for example, were reconstructed. 

        However, in cases where Fourier analysis is not possible or reliable (sparse sampling, highly variable viewing geometry, too few data points), we can use a brute-force approach to find the best model by testing all physically possible combinations of parameters. By starting the inversion algorithm from many initial parameter values, we can scan the whole physically acceptable parameter space and find the best set of parameters. The measure of goodness-of-fit is the usual $\chi^2$. This approach is time-consuming yet robust as we not only find the best solution but also show that all other combinations of spin parameters are not compatible with data. In practice, however, many models with different parameters usually fit the data well (low $\chi^2$). This means that the spin state cannot be determined uniquely. However, for many of these formally good models, the shape-related moments of inertia $I_1^\mathrm{shape}$ and $I_2^\mathrm{shape}$ do not correspond with the fitted values $I_1$ and $I_2$. Such models are self-inconsistent, and they have to be rejected. This makes the whole inversion process slow and complicated because each acceptable model has to be converted from its Gaussian image to vertex representation, and the consistency of inertia tensors has to be checked.
  
        To avoid this possible inconsistency in the model and make the scanning of the parameter space faster, we used an alternative approach. Instead of a convex shape model, we used a triaxial ellipsoid. It is a much simpler model but can reproduce the primary light curve features well. Its advantage is that the kinematic parameters are coupled with the shape, and the model is self-consistent. Another significant advantage is that the disk-integrated brightness of a triaxial ellipsoid can be computed analytically when assuming that scattering is geometric \citep{Con.Ost:84}, which makes computation much faster than in the case of convex models. 
  
        The shape is approximated with a triaxial ellipsoid with semiaxes $a > b \geq c = 1$. The model's orientation $\phi_0$, $\psi_0$, and direction of the angular momentum vector $\lambda$, $\beta$ are described by the same angles as in the case of the convex model. Principal moments of inertia tensor can be easily computed from semiaxes $a$, $b$, and vice versa ($c = 1$, $I_3 = 1$):
        
        \begin{equation}
            I_1 = \frac{b^2 + 1}{a^2 + b^2}, \qquad I_2 = \frac{a^2 + 1}{a^2 + b^2}\,.
        \end{equation}
        In this approach, parameters $I_1$ and $I_2$ simultaneously describe the kinematics of the rotation and shape.

        \begin{figure}[t]
            \includegraphics[width=\columnwidth]{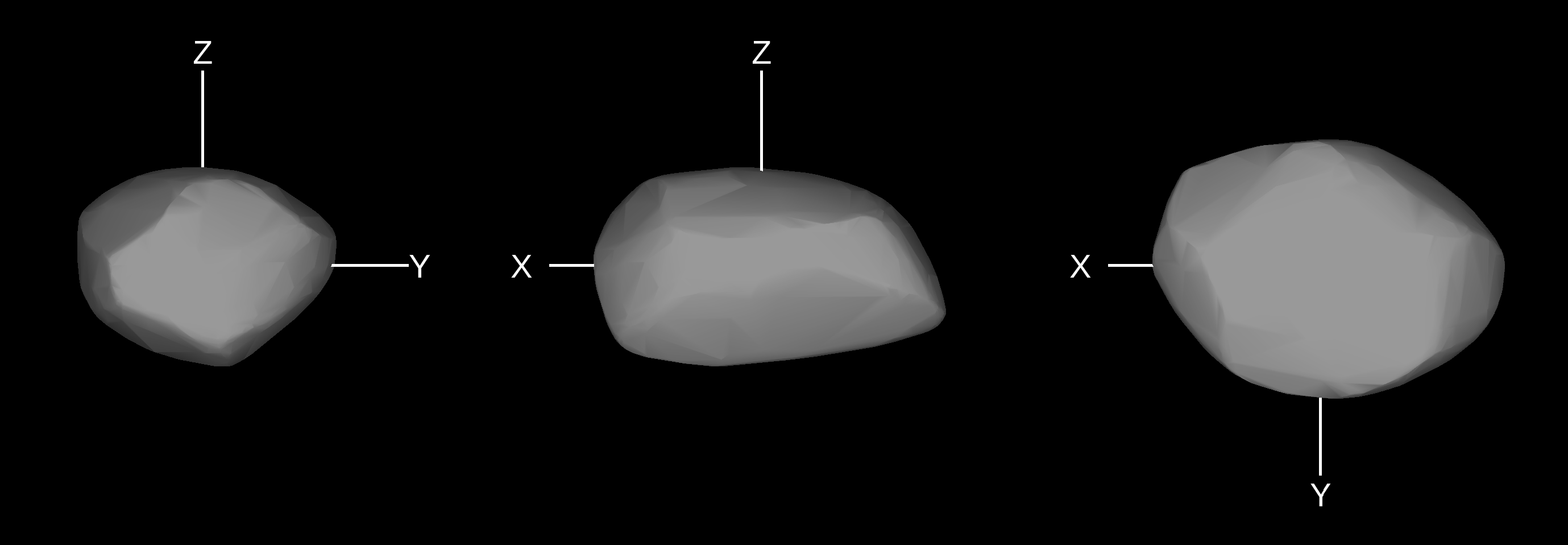}
            \includegraphics[width=\columnwidth]{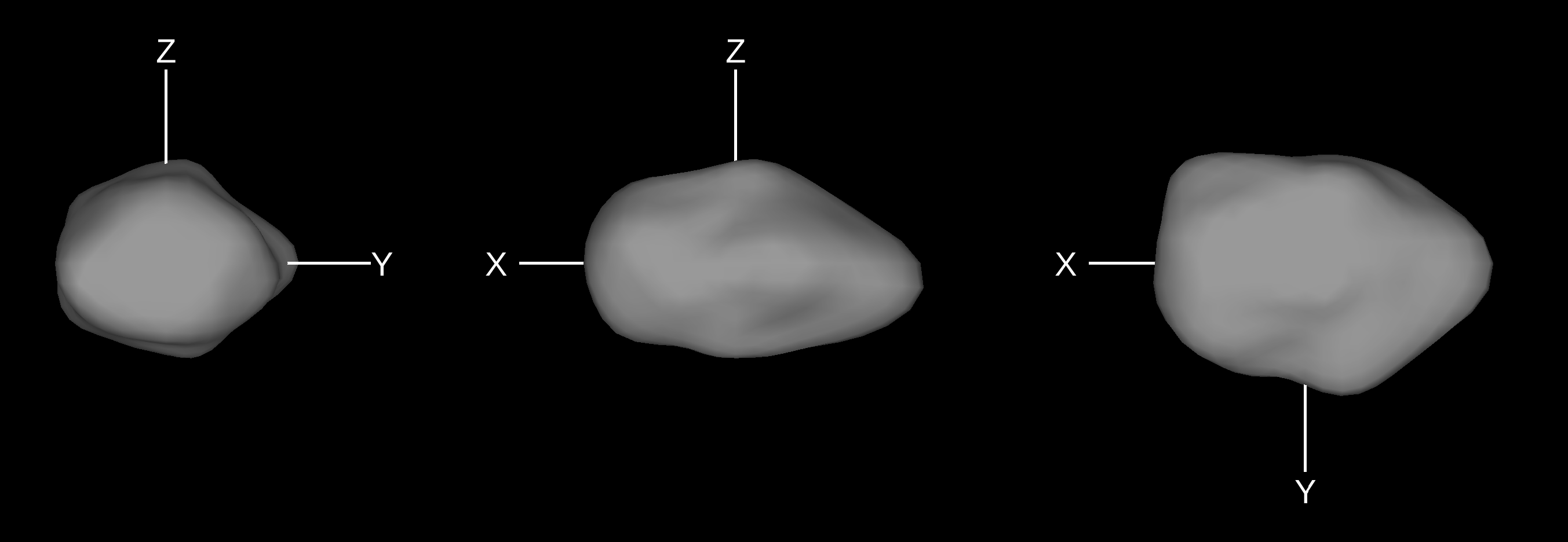}
            \caption{Shape models of Leona. The convex (top) and nonconvex (bottom) shape models of (319)~Leona are shown from three directions. The convex model was reconstructed from light curves only, and the nonconvex one from light curves and occultations.}
            \label{fig:shape}
        \end{figure}

    \subsection{Finding the rotation parameters with the ellipsoidal model}

        The optimization algorithm of \cite{Kaa:01} is gradient-based, so it robustly converges to the local minimum of $\chi^2$. To find the global minimum, the optimization has to be started from different initial values to cover all local minima. The critical part is to find the correct values of $P_\phi$ and $P_\psi$. The separation between local minima depends on the length of the time interval $T$ covered by observations. A rough estimate can be computed similarly to a single-period light curve as $0.5P^2/T$, where $P$ is the period \citep{Kaa:04}. 

        Instead of scanning the entire parameter space using the whole photometric data set, we first used only a subset covering a smaller time interval $T$. Thus, the computation was faster because there were fewer local minima, and the number of data points was smaller. This first step dramatically reduced the size of the parameter space with acceptable solutions. Then, we added more data, increased $T$, scanned the remaining intervals of parameters more densely, and found the final set of parameters.
        
        The first data set we used was that of \cite{Pil.ea:17b}. We obtained the best solution with the precession period $P_\phi = 319$\,h and the rotation period $P_\psi = 1210$\,h. The main period of light curve data reported by \cite{Pil.ea:17b} was $\sim 431$\,h, which corresponds to the difference in physical frequencies $1/(1/P_\phi - 1/P_\psi) = 433$\,h. When this best-fit solution was localized, we added the remaining photometric data described in Sect.~\ref{sec:new_observations} and searched on a fine grid around these values. This enabled us to determine the periods precisely to $P_\phi = 314.23$\,h and $P_\psi = 1172.0$\,h. The best-fit ellipsoid had semiaxes $a = 1.97$, $b = 1.35$, corresponding to $I_1 = 0.495$, $I_2 = 0.857$. Then, we used the parameters from the ellipsoidal fit as initial parameters for the convex inversion method with the complete data set. 
        
        \begin{figure*}[t]
            \includegraphics[width=\textwidth]{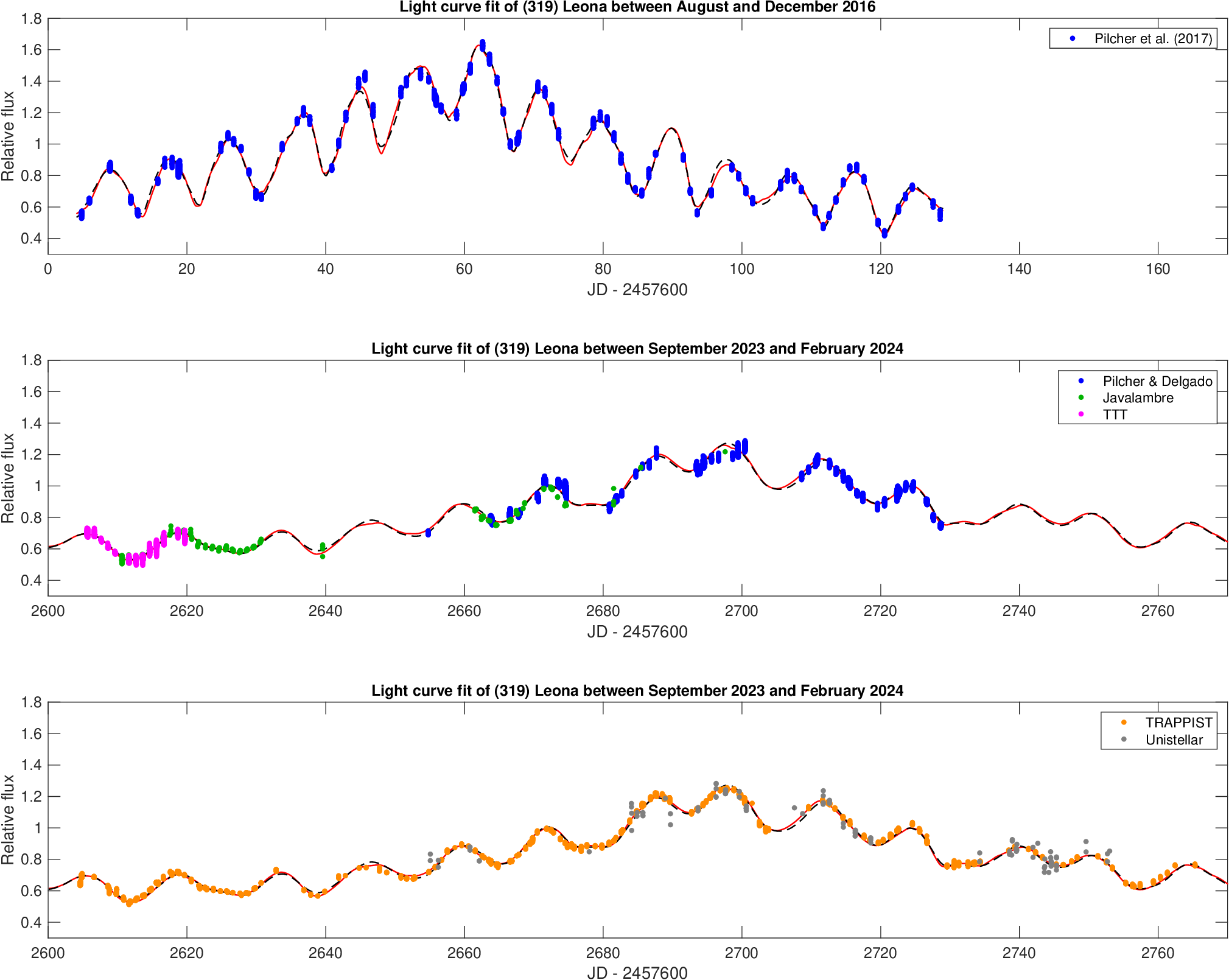}
            \caption{Comparison between model light curves and observations. The light curves predicted by the convex shape model (solid red curve) and the nonconvex shape model (dashed black curve) are plotted against the observed photometric data (points). The top panel shows data by \cite{Pil.ea:17b} from 2016; the middle and bottom panels show five sets of new observations from 2023/24.}
            \label{fig:lcfit}
        \end{figure*}

        \begin{figure*}[t]
            \includegraphics[width=\textwidth]{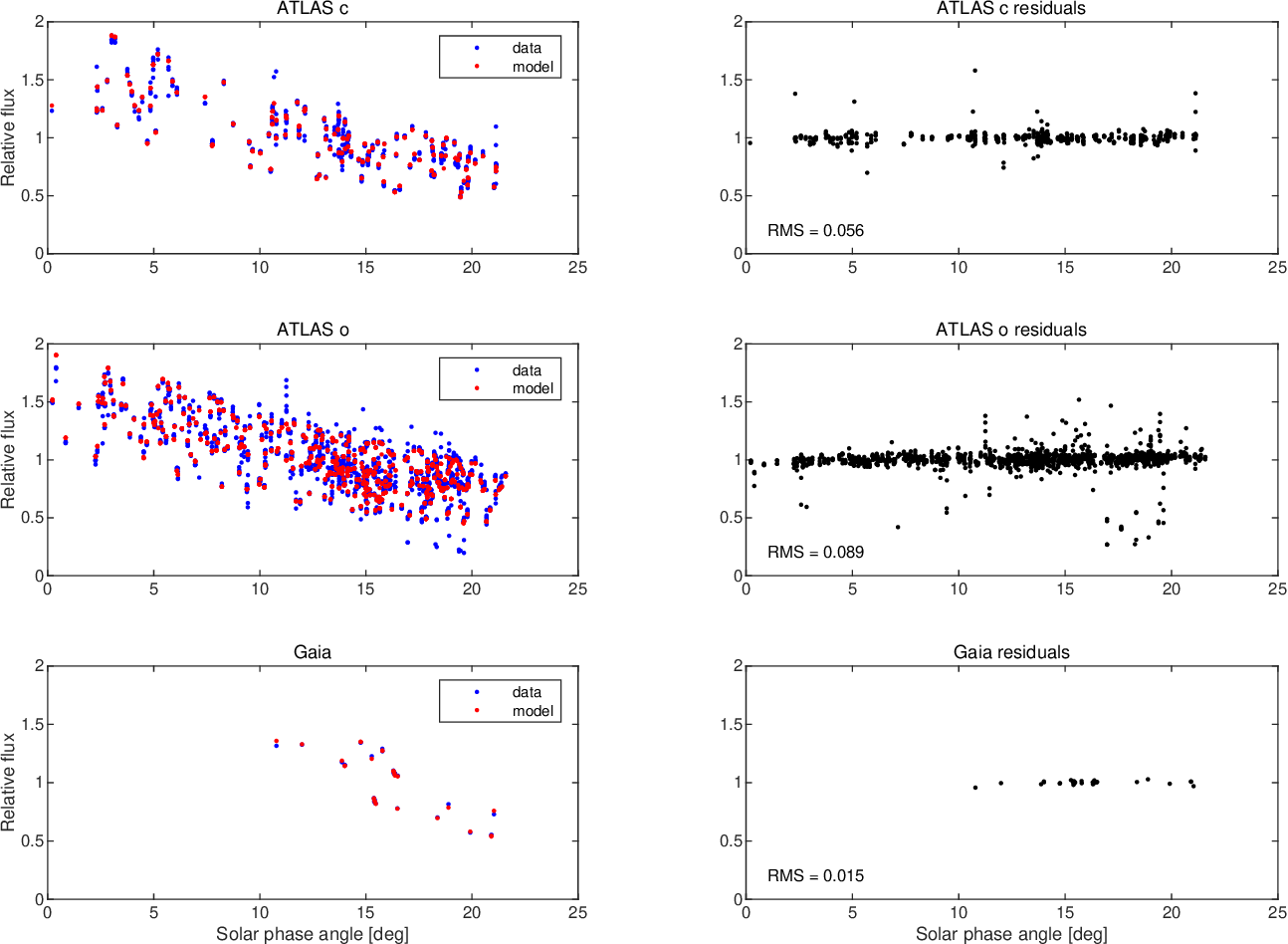}
            \caption{Sparse photometry from surveys. The left panels show the comparison between the light curve predicted by the convex shape model (red points) and observed photometric data (blue points) from ATLAS in c and o filters and Gaia. The brightness in relative flux units is plotted against the solar phase angle to see the phase curve. The three panels on the right show the difference between observed and modeled brightness and the values of root mean square residuals.}
            \label{fig:lcfit_sparse}
        \end{figure*}
       
    \subsection{Convex shape model}
    \label{sec:convex}

        When applying the convex inversion method to the complete data set using the best parameters from the ellipsoidal model as initial values, we found two possible shape models that differed in the direction of the angular momentum vector. Still, only one was consistent with the occultations (see Sect.~\ref{sec:occ}). The parameters of this best-fit model are listed in Table~\ref{tab:spin}, the convex shape model is shown in Fig.~\ref{fig:shape}, and the fit to observations is shown in Figs.~\ref{fig:lcfit} and \ref{fig:lcfit_sparse}. The 3D shape model and its spin parameters are available online in the Database of Asteroid Models from Inversion Techniques.\footnote{\url{https://astro.troja.mff.cuni.cz/projects/damit/}}

        To estimate the uncertainty of the spin parameters, we created one hundred bootstrap samples of the data by randomly selecting samples of the original data. We used only the three main data sets -- (i) photometry from 2016 by \cite{Pil.ea:17b}, (ii) Pilcher's and Delgado's data from 2023, and (iii) TRAPPIST data from 2023/24. So, each bootstrapped sample consisted of these three light curves with randomly sampled data points. We repeated the inversion for five initial pairs of $I_1$ (0.46, 0.47, 0.48, 0.495, 0.51) and $I_2$ (0.865, 0.855, 0.855, 0.845, 0.835), so we had 500 bootstrap models. The uncertainties reported in Table~\ref{tab:spin} are standard deviations of the bootstrap parameters. The uncertainty of inertia values is large, with a strong negative correlation between $I_1$ and $I_2$. The best combination is $I_1 = 0.495$ and $I_2 = 0.845$, but other acceptable solutions concentrate around values $I_1 = 0.460$ and $I_2 = 0.865$. The values computed from the shape model with uncertainties estimated from bootstrap are $I_1^\mathrm{shape} = 0.54 \pm 0.01$ and $I_2^\mathrm{shape} = 0.87 \pm 0.01$, which is sufficiently close to the kinematic moments of inertia given the systematic uncertainty of the convex shape model. For example, stretching the shape by 10\% along the $x$ axis and 7\% along the $y$ axis changes the values to $I_1^\mathrm{shape} = 0.493$ and $I_2^\mathrm{shape} = 0.849$, which is practically the same as the kinematic values $I_1, I_2$. When scaled by occultation, the diameter of this modified shape is 58.5\,km.

        We used Hapke's light-scattering model \citep{Hap:12} with a fixed value of surface roughness slope $\bar\theta = 20^\circ$ and optimized the single-scattering albedo $w$, opposition-surge parameters $h$ and $B_0$, and the asymmetry factor $g$. The best-fit values are listed in Table~\ref{tab:spin}. However, because we did not use the information about the size of the asteroids at this stage of modeling, Hapkes's parameters are only poorly constrained. The set presented in Table~\ref{tab:spin} is just one of many possible combinations. The geometric albedo, which is uniquely defined by Hapke's parameters, is $p = 0.07$, which is consistent with the range 0.02--0.08 compiled at The Minor Planet Physical Properties Catalogue (MP3C).\footnote{\url{https://mp3c.oca.eu}}

    \begin{figure*}[t]
        \includegraphics[width=0.49\textwidth]{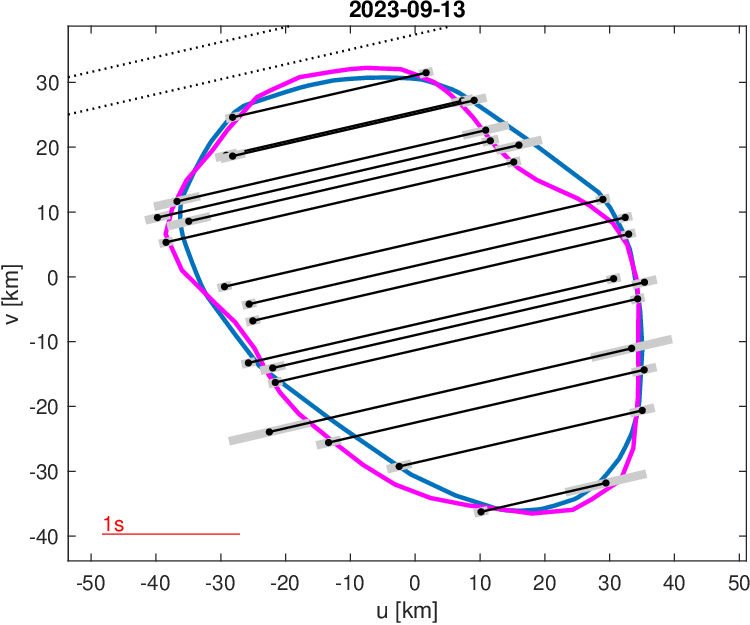}
        \includegraphics[width=0.49\textwidth]{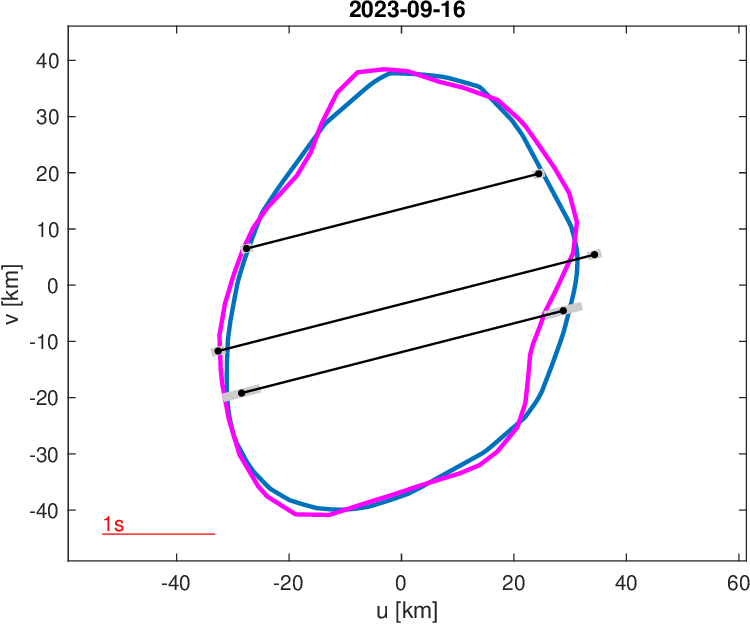}\\
        \includegraphics[width=0.49\textwidth]{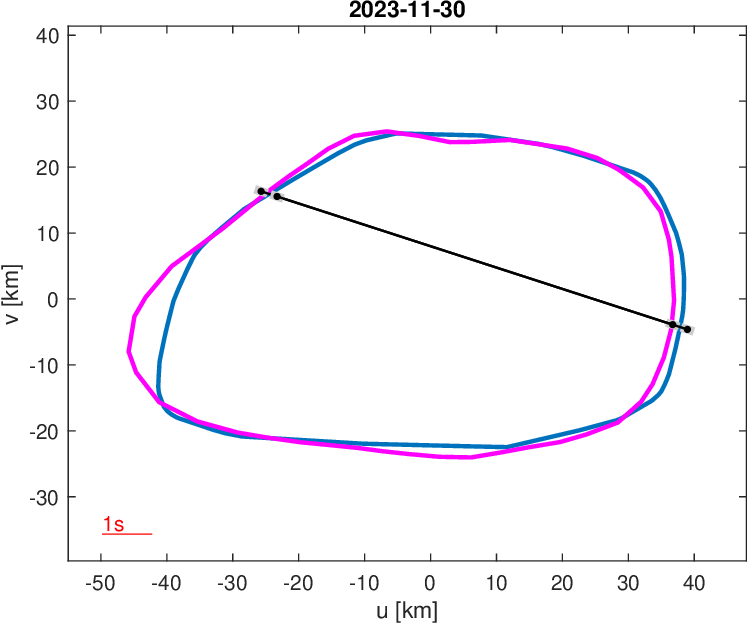}
        \includegraphics[width=0.49\textwidth]{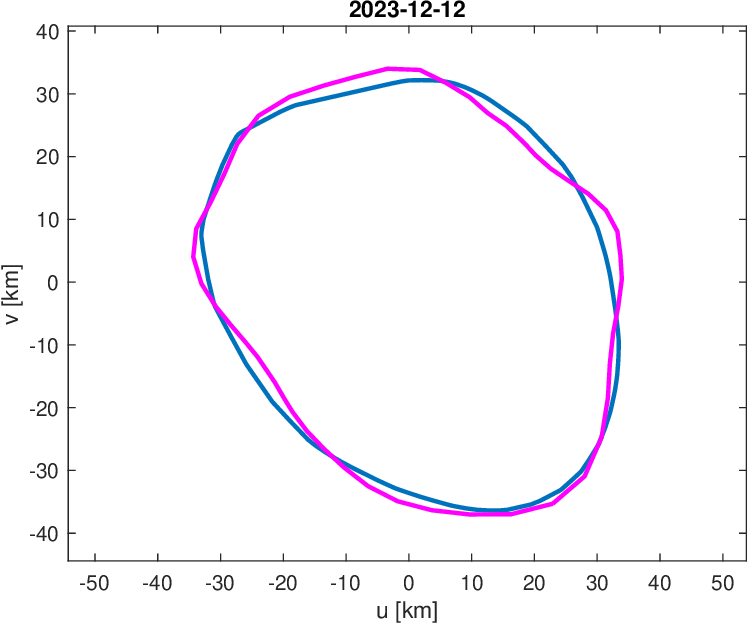}
        \caption{Projections of the models. Silhouettes of the best-fit convex (blue) and nonconvex (magenta) shape models are shown for four occultations. The chords are shown as solid lines with gray timing error bars at their ends. The dotted straight lines in the first panel are negative observations. The red line gives the time scale, which is one second long. The last panel shows the prediction of the silhouette for the Betelgeuse occultation.}
        \label{fig:occ}
    \end{figure*}

    \section{Scaling the model by occultations}
    \label{sec:occ}

        \subsection{Scaling the convex model}

        The convex shape model derived above is scale-free; that is, its size is not known. Although we used Hapke's photometric model to fit calibrated photometry, absolute calibration was not considered. The light curve sets were used as relative light curves that can be arbitrarily shifted on the magnitude scale. So, we can now compute the model's orientation for the time of Betelgeuse orientation, but the size of the asteroid's projection is still unknown.

        To scale the model, we used three stellar occultations by Leona observed in 2023 and summarized in Table~\ref{tab:occ}. The data were extracted from the Occult database, which is available online.\footnote{\url{https://www.occultations.org/sw/occult/Asteroid_Observations.zip}} The size of Leona was found the same way as in \cite{Dur.ea:11}. The model was projected on the fundamental plane for the times of occultations, and its size was optimized such that the match between the silhouettes and the chords was at best. The first two occultations from September 2023 constrain the size well. The third from November 2023 consists of only two chords that lie on each other, so they are consistent with almost any silhouette, as the relative shift between the center of the projection and the chords is a free parameter. Relative shifts larger than the ephemeris uncertainties might be, in principle, forbidden, but problems with the occulted star position and other scenarios can cause real discrepancies, so this potential constraint has not been implemented in the optimization algorithm. In Fig.~\ref{fig:occ}, we show projections of the scaled shape model for those three occultations and also for the time of Betelgeuse occultation.

\begin{figure}[t]
        \includegraphics[width=\columnwidth]{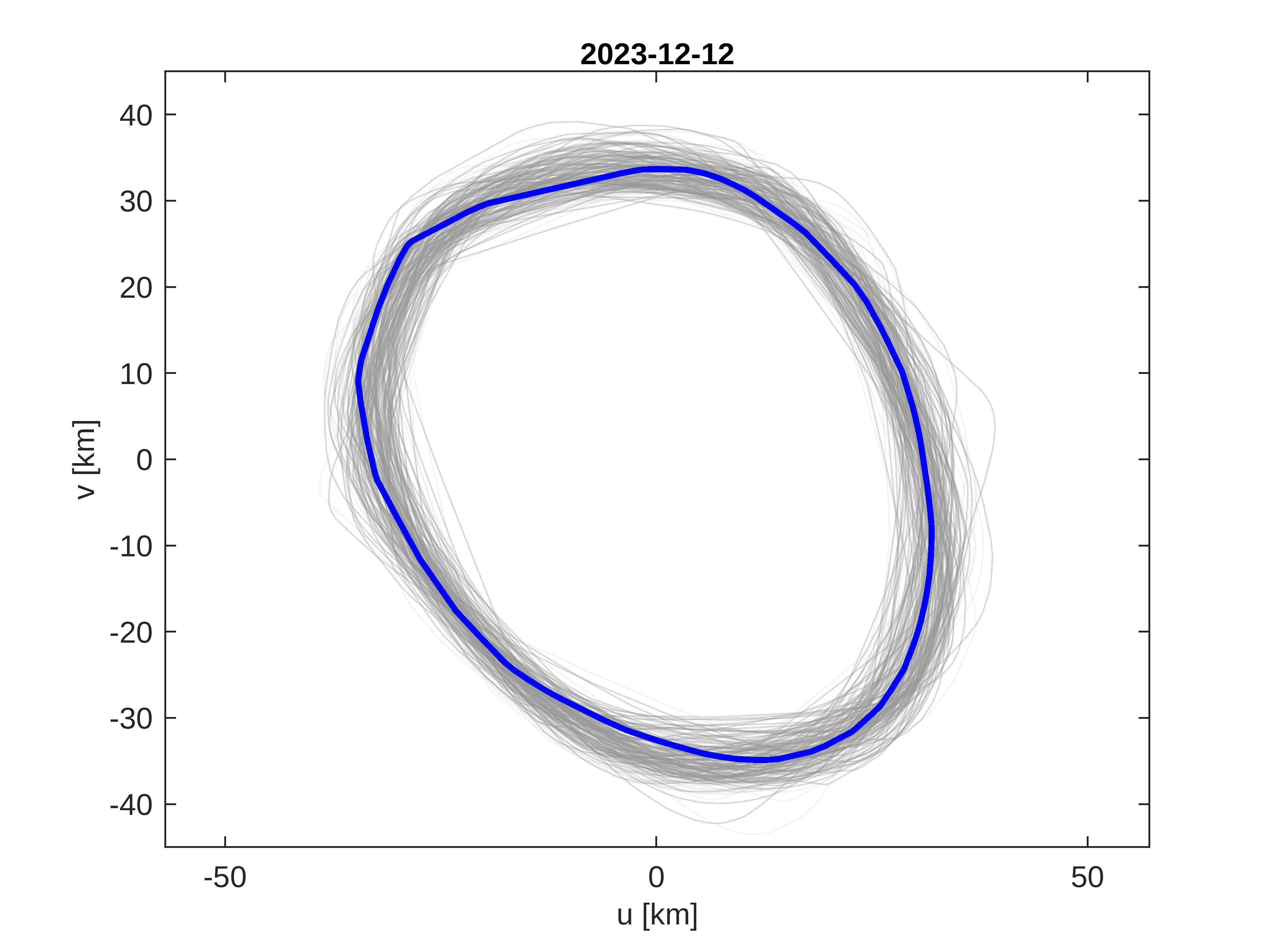}
        \caption{Projections of 500 bootstrap shape models for the time of Betelgeuse occultation. The blue contour corresponds to the best-fit model.}
        \label{fig:bs_projections}
    \end{figure}

        The volume-equivalent diameter of Leona's convex shape model obtained this way was 59.1\,km. From the bootstrap sample, the standard deviation is 0.9\,km (mean value 59.1\,km), which we take as a realistic $1\sigma$ uncertainty of the size of the convex model. To visualize the uncertainty of our prediction, we show in Fig.~\ref{fig:bs_projections} projections of all 500 bootstrap models for the time of Betelgeuse occultation. Although the formal uncertainty of the equivalent size is only $\sim 1$\,km, the shapes and the spin parameters cause the projections to fluctuate around the best-fit model substantially.

  \subsection{Nonconvex model}
  \label{sec:nonconvex}

    The convex model derived above describes Leona's global shape. However, real shapes of asteroids are never mathematically convex, so the convex approximation will always be just an approximation. When many chords cover a 2D projection of an asteroid during an occultation, nonconvex features can be directly seen. The first occultation by Leona from 13 September 2023 provided considerable detail on the body thanks to the 25 positive chords \citep{Ort.ea:24} and indeed shows some inconsistencies between the convex model and the observed chords (Fig.~\ref{fig:occ}). These inconsistencies can be interpreted as concavities of the real shape and can be improved by creating a nonconvex shape model.

    To be able to model a nonconvex shape of Leona, we modified the Knitted Occultation, Adaptive-optics, and Lightcurve Analysis (KOALA) algorithm of \cite{Car.ea:12} to work with tumbling asteroids. Similarly, like in the case of bootstrap error estimation, we used only the three best data sets. Vertices and surface triangles represent the shape model; the optimization is done simultaneously concerning the light curves and occultations. It is necessary to include some regularization of the shape to avoid shapes that are too irregular and have spiky surfaces. So, the best-fit shape depends on the setup of the regularization and the weight between the light curve and occultation part. As an example, we show in Fig.~\ref{fig:shape} a realistically looking nonconvex model that fits the light curves at about the same level as the convex model (Fig.~\ref{fig:lcfit}) and fits the occultation data better than a convex model (Fig.~\ref{fig:occ}). Its volume-equivalent diameter is 59.5\,km. For comparison, the MP3C lists 15 published diameter values ranging from 50 to 89\,km with the mean value around 64\,km. The kinematic principal moments of inertia are the same as for the convex model (Table~\ref{tab:spin}), while the moments of inertia computed from the shape are $I_1^\mathrm{shape} = 0.525$ and $I_2^\mathrm{shape} = 0.889$.

  \section{Conclusions and future prospects} 

    Using asteroid Leona's rotational light curves and sparse photometry, together with observed stellar occultations, we reconstructed its spin state and shape model. This enabled us to predict Leona's orientation when it occulted Betelgeuse on 12 December 2023. The profile of Leona for the Betelgeuse occultation is provided as an ASCII file in online material. The projection of Leona is necessary to analyze light curves recorded during the occultation and reconstruct the brightness distribution across the disk of Betelgeuse. However, this complex task is beyond the scope of this paper. 

    If recorded by many observers to densely cover the projected silhouette, future stellar occultations by Leona could further constrain its nonconvex shape, which would potentially improve the analysis of the Betelgeuse occultation. There will be several occultation events in the 2025 to 2026 time frame with good potential to provide multichord observations. One of them, with good visibility in the USA, will happen on 18 April 2025. All the relevant details and an interactive map is available online\footnote{\url{https://astro.kretlow.de/cora/occultations/60eb1840-df12-11ef-7f69-edd2bbd75275/}}. Observations can be coordinated at the Occultation Portal\footnote{\url{https://occultationportal.org/chords/2198}}. Other events by Leona can be directly retrieved online at the CORA site\footnote{\url{https://astro.kretlow.de/cora}} by selecting the special occultation database and selecting Leona in the target name field.

  \section{Data availability}

    The projections of the convex and nonconvex models for the time of Betelgeuse occultation (Fig.~\ref{fig:occ}, bottom right panel) are available as data files at \url{https://doi.org/10.5281/zenodo.15059639}.
    
  \begin{acknowledgements}
    This work has been supported by the grant 22-17783S of the Czech Science Foundation. This publication makes use of data products from the TRAPPIST project. TRAPPIST-South is funded by the Belgian National Fund for Scientific Research (F.R.S.-FNRS) under grant PDR T.0120.21. TRAPPIST-North is funded by the University of Liège and performed in collaboration with Cadi Ayyad University of Marrakech. EJ is Director of Research at Belgian FNRS. Part of this work was supported by the Spanish projects PID2020- 112789GB-I00 from AEI and Proyecto de Excelencia de la Junta de Andalucía PY20-01309. Financial support from the grant CEX2021- 001131-S funded by MCIN/AEI/ 10.13039/501100011033 is also acknowledged. This article includes observations made in the Two-meter Twin Telescope (TTT) sited at the Teide Observatory of the Instituto de Astrof\'isica de Canarias (IAC), that Light Bridges operates in Tenerife, Canary Islands (Spain). The Observing Time Rights (DTO) used for this research were provided by RICTEL TTT, SA. MSR used storage and computing capacity in ASTRO POC's EDGE computing center in Tenerife under the form of Indefeasible Computer Rights (ICR). The ICR were provided by Light Bridges, SL with the collaboration of Hewlett Packard Enterprise and VAST DATA. This work is also partly based on observations made with the Tx40 telescope at the Observatorio Astrofísico de Javalambre in Teruel, a Spanish Infraestructura Cientifico-Técnica Singular (ICTS) owned, managed, and operated by the Centro de Estudios de Física del Cosmos de Aragón (CEFCA). Tx40 is funded by the Fondos de Inversiones de Teruel (FITE). The work of AM was supported by the National Science Centre, Poland, through grant no. 2020/39/O/ST9/00713. This work has made use of data from the European Space Agency (ESA) mission {\it Gaia} (\url{https://www.cosmos.esa.int/gaia}), processed by the {\it Gaia} Data Processing and Analysis Consortium (DPAC, \url{https://www.cosmos.esa.int/web/gaia/dpac/consortium}). Funding for the DPAC has been provided by national institutions, in particular, the institutions participating in the {\it Gaia} Multilateral Agreement. This work has made use of data from the Asteroid Terrestrial-impact Last Alert System (ATLAS) project. ATLAS is primarily funded to search for near-Earth asteroids through NASA grants NN12AR55G, 80NSSC18K0284, and 80NSSC18K1575; byproducts of the NEO search include images and catalogs from the survey area. The ATLAS science products have been made possible through the contributions of the University of Hawaii Institute for Astronomy, the Queen's University Belfast, the Space Telescope Science Institute, and the South African Astronomical Observatory. This work is based on data provided by the Minor Planet Physical Properties Catalogue (MP3C) of the Observatoire de la Côte d'Azur.
  \end{acknowledgements}

  \bibliographystyle{aa}
  \bibliography{bibliography_all,mybib}

  \begin{appendix}

    \section{Unistellar observers and occultation observers}
    
    \begin{table}[h]
        \caption{List of names of Unistellar observers that waived anonymity.}
            \label{tab:unistellar}
            \begin{tabular}{lll}
            \hline
J. Archer & 
R. Blake &
N. Delaunoy \\
K. Fukui &
A. Gillmartin &
P. Girard \\
T. Goto &
B. Guillet &
B. Haremza \\
D. Havell &
P. Heafner &
P. Huth \\
R. Knight &
D. Koster &
R. Kukita \\
P. Kuossari &
J. Laugier &
S. Lawrence \\
C. Logan &
Y. Lorand &
N. Meneghelli \\
M. Mitchell &
F. Mortecrette &
W. Ono \\
J. Pickering &
M. Primm &
F. Ribas \\
D. Rivett &
M. Shimizu &
K. Sibbernsen \\
G. Simard &
S. Stahl &
B. Tobias \\
C. Vantaggiato &
S. Will &
P. Yehle \\
N. Yoblonsky &
W. Yue &
K. Zajdel \\
\hline
            \end{tabular}

        \end{table}
  
   \begin{table}[h]
        \caption{List of observers, and their locations, who participated in three occultations by Leona shown in Fig.~\ref{fig:occ}.}
            \begin{tabular}{l}
            \hline
            \multicolumn{1}{c}{13 September 2023} \\[1mm]
            V.~Dekert \& N.~Morales, Observatoire De La Sagra, Spain \\
            A.~Castillo, Donadio, Jaen, Spain \\
            R.~Goncalves, Beja, Portugal \\
            J.~Rovira, Casas De Don Juan, Huescar, Spain \\
            M.~Sanchez \& S.~Moral, Jodar, Jaen, Spain \\
            J.~Marti \& P.~L.~Luque Escamilla, Universidad De Jaen, Spain \\
            C.~Perello \& C.~Schnabel, Pozo Alcon, Spain \\
            V.~Pelenjow, Castillejar, Spain \\
            A.~Roman \& S.~Alonso, Pegalajar, Jaen, Spain \\
            J.~Flores, Huelma, Spain \\
            J.~M.~Fernandez, Cerro Negro, Sevilla, Spain \\
            A.~Selva \& C.~Schnabel, Mirador Del Negratin, Spain \\
            R.~G.~Farfan, Ecija, Spain \\
            N.~Morales \& J.~L.~Rizos, Campotejar, Spain \\
            F.~Casarramona \& E.~Smith, Los Coloraos, Gorafe, Spain \\
            J.~L.~Maestre, Albox, Spain \\
            J.~Delgado, Camas, Sevilla, Spain \\
            J.~M.~Madiedo, Sevilla, Spain \\
            J.~Flores, Calar Alto Observatory, Spain \\
            \\
            \multicolumn{1}{c}{16 September 2023} \\[1mm]
            R.~Venable, Holbrook, AZ, USA \\
            R.~Venable, Heber, AZ, USA, 2 stations \\
            \\
            \multicolumn{1}{c}{30 November 2023} \\[1mm]
            K.-L. Bath \& D. Husar, Hakos, Namibia \\
            S.~Meister \& P.~Englmaier, Hakos, Namibia \\
            \hline
            \end{tabular}
            \label{tab:occ}
        \end{table}

  \end{appendix}
      
\end{document}